\documentclass[aps,prl,reprint,superscriptaddress]{revtex4-1}
\usepackage{H1}
\usepackage{amsmath}
\usepackage[pdftex,colorlinks=true]{hyperref}
\hypersetup{
citecolor = blue
}
\usepackage[normalem]{ulem}
\usepackage{amssymb}
\usepackage{amsthm}
\usepackage{amsfonts}
\usepackage{bbm}
\usepackage{array}

\begin{document}
\title{Two universality classes for the many-body localization transition}

\author{Vedika Khemani}
\affiliation{\mbox{Department of Physics, Harvard University, Cambridge, MA 02138, USA}}
\author{D. N. Sheng}
\affiliation{\mbox{Department of Physics and Astronomy, California State University, Northridge, CA 91330, USA}}
\author{David A. Huse}
\affiliation{\mbox{Department of Physics, Princeton University, Princeton, NJ 08544, USA}}

\begin{abstract}
We provide a systematic comparison of the many-body localization transition in spin chains with nonrandom quasiperiodic vs. random fields.  We find evidence suggesting that these belong to two separate universality classes: the first dominated by ``intrinsic'' \emph{intra}-sample randomness, and the second dominated by external \emph{inter}-sample quenched randomness. We show that the effects of inter-sample quenched randomness are strongly growing, but not yet dominant, at the system sizes probed by exact-diagonalization studies on random models. Thus, the observed  finite-size critical scaling collapses in such studies appear to be in a preasymptotic regime near the nonrandom universality class, but showing signs of the initial crossover towards the external-randomness-dominated universality class. Our results provide an explanation for why exact-diagonalization studies on random models both see an apparent scaling near the transition while also obtaining  finite-size scaling exponents that strongly violate Harris/Chayes bounds that apply to disorder-driven transitions.   We also show that the MBL phase is more stable for the quasiperiodic model as compared to the random one, and the transition in the quasiperiodic model suffers less from certain finite-size effects. 
\end{abstract}

\maketitle

Many-body localization (MBL) generalizes the phenomenon of Anderson localization 
to the interacting setting \cite{Anderson58, Basko06, PalHuse, OganesyanHuse, Nandkishore14, AltmanVosk}. The dynamics in an MBL system fails to establish local thermal equilibrium, and even highly excited states can retain local memory of their initial conditions for arbitrarily late times. The transition between an MBL phase and a ``thermalizing'' one is not a thermodynamic phase transition and lies outside the framework of equilibrium statistical mechanics. Instead it is a novel \emph{eigenstate phase transition} \cite{Huse13, PekkerHilbertGlass}  across which thermal and ``volume-law'' entangled many-body eigenstates obeying the eigenstate thermalization hypothesis (ETH) \cite{Deutsch, Srednicki, Rigol}  change in a singular way to non-thermal and area-law entangled eigenstates in the MBL phase.

Although the MBL transition has attracted much recent interest~\cite{Kjall14,Luitz15,GroverCP,SerbynCriterion,AgarwalGriffiths2015,Devakul15,santos, mobilityedge,de2016stability,SerbynSpectralStats,GopalakrishnanGriff, ZhangFloq,ClarkBimodal,KhemaniCP,VHA,PVP,ZhangRG,DVP}, very little is definitively known about its properties. Phenomenological renormalization group (RG) treatments of the transition are approximate but can probe large system sizes, and such studies \cite{VHA,PVP,ZhangRG, DVP} find a continuous transition in one dimension with a finite-size critical scaling exponent $\nu_{FS}  \sim 3$ satisfying rigorous Harris/CCFS/CLO scaling bounds \cite{Harris, CCFS2, CLO} which require $\nu_{FS}  \geq 2/d$ for transitions in $d$ dimensions in the presence of quenched randomness.
On the other hand, most other studies of the transition use numerical exact diagonalization (ED) of spin-chains which is limited to system sizes $L \leq 22$.  These ED studies observe an apparent scaling collapse near the transition, but with scaling exponents $\nu_{FS}  \sim 1$  violating the CCFS/CLO bound \cite{Kjall14, Luitz15}. Strikingly, some aspects of this transition even look first-order-like in that quantities like the eigenstate entanglement entropy (EE) of small subsystems can vary discontinuously across the transition~\cite{KhemaniCP, DVP}.

\begin{figure}
  \includegraphics[width=\columnwidth]{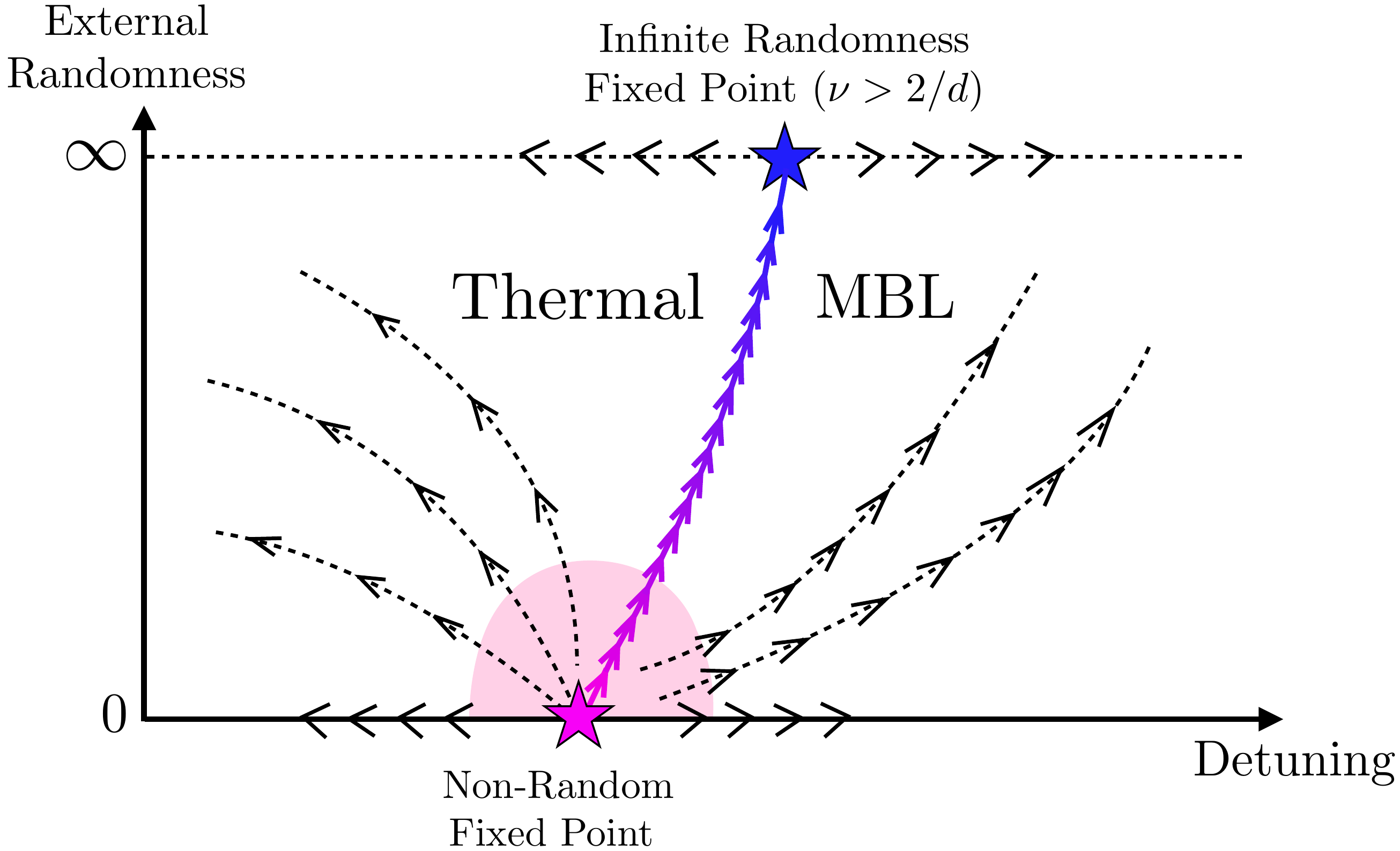}
  \caption{\label{fig:RG} Schematic RG flow for a one dimensional system displaying an MBL transition. In the absence of external randomness, the critical fixed point is dominated by ``intrinsic'' \emph{intra}-sample variations and is not constrained by Harris/Chayes bounds (pink star). The addition of external quenched randomness is a Harris-relevant perturbation which causes the nonrandom fixed point to flow towards an ``infinite randomness'' disorder dominated fixed point (blue star). The ``detuning'' parameter quantifies the ratio of off-diagonal to diagonal couplings in the most local basis for the coarse grained model. The MBL phase is more stable in the nonrandom model and thus the 
critical flow is towards higher detuning. We propose that the effects of external randomness are not yet fully apparent at the sizes probed by ED studies, and the transition in these systems is mostly still governed by the nonrandom fixed point while beginning to crossover towards the random fixed point (shaded oval).     }
\end{figure}

A sensitive probe of the MBL transition is the standard deviation of the half-chain EE, $\Delta_S$, which peaks at the transition as the eigenstates change from area law to volume law entangled~\cite{Kjall14}. A careful parsing of $\Delta_S$ across inter- and intra- sample contributions near the transition reveals two notable features~\cite{KhemaniCP}: (i)  a sizeable volume-law scaling for $\Delta_S$ across eigenstates of the \emph{same} sample, a property that none of the RG treatments capture and, (ii) a super-linear growth with $L$ for the sample-to-sample contribution to $\Delta_S$ at the system sizes studied by ED, a trend that is unsustainable in the large $L$ limit since the maximum possible EE scales as a volume law. This parsing indicates that the observed violations of CCFS/CLO bounds (which are derived from sample-to-sample variations) might result from a scenario in which the effect of quenched randomness across samples is not yet fully manifest, but growing strongly, at the sizes probed by ED~\cite{KhemaniCP}. These data also suggest an intriguing scenario in which there might be two universality classes for transitions between MBL and thermal phases: one dominated by ``intrinsic'' eigenstate randomness within a given sample, and the second dominated by external quenched randomness across samples. In this scenario, the observed critical finite-size scaling collapses would appear to be in a preasymptotic regime near the first universality class (for which CCFS/CLO bounds do not apply), but showing the signs of the initial crossover towards the second external-randomness dominated universality class.

In this letter, we provide a more systematic analysis of the scenario above by studying the MBL transition in a quasiperiodic (QP) model with no quenched randomness. Following the work of Aubry and Andr\'{e}~\cite{aubry1980}, the localization transition in non-interacting quasiperiodic models has been extensively studied. More recently, it was shown that interacting quasiperiodic models have an MBL phase \cite{IyerQP}, and signatures of this phase have been observed in cold-atomic experiments \cite{Schreiber2015,Bordia2016}.  However, compared to its non-interacting counterpart, the  MBL transition in quasiperiodic models has received little theoretical attention.  Nor have the points of similarity and difference between the MBL transition in quasiperiodic and random models been systematically studied. In this work, we provide a detailed finite-size scaling analysis of the QP-MBL transition, along with a comparison to the random MBL transition.  We find that the MBL phase is more stable for the quasiperiodic model than for the random one, which is opposite to the trend for single-particle localization.  This we attribute to the effects of locally thermal rare regions that destabilize MBL in the random system.  The finite-size scaling we find suggests that there is a nonrandom universality class of the transition, and \emph{both}  models are governed by this universality class for the sizes accessible to ED.  However, the random model is beginning to cross over towards the external-randomness-dominated universality class. Adding randomness to the quasiperiodic model is thus a ``Harris-relevant'' perturbation, causing this crossover. Fig~\ref{fig:RG} shows a schematic RG flow for this scenario. Altogether, our work not only advances our understanding of the global structure of quantum criticality in MBL systems, but also provides a concrete explanation for why numerical studies on random models see finite-size scaling collapse but obtain exponents violating Harris/Chayes bounds.

\noindent{\bf Model:}
We consider quasiperiodic/random spin chains of the form
\begin{align}
H^{QP/R} &= J\sum_{i=1}^{L-1} (S_i^xS_{i+1}^x + S_i^y S_{i+1}^y )  +  J_z \sum_{i=1}^{L-1} S_i^zS_{i+1}^z \nonumber\\
&+ \sum_{i=1}^{L} W\cos(2 \pi k i +\phi_i^{QP/R} ) S_i^z \nonumber\\
& + J'\sum_{i=1}^{L-2} (S_i^xS_{i+2}^x +S_i^yS_{i+2}^y )
\label{eq:model}
\end{align}
where $S_i^{\{x/y/z\}}$ are spin 1/2 degrees of freedom on site $i$,  $J=J'=J_z=1$, and $k= \frac{\sqrt{5}-1}{2}$ is an irrational wavenumber. For the quasiperiodic model, $\phi_i^{QP} = \phi \in [-\pi, \pi)$ is an arbitrary global phase offset such that the on-site fields are periodic with a period that is incommensurate with the lattice. This choice with  $J' = J_z =0$ is the non-interacting Aubry-Andr\'{e} model which is localized for $ W > 1$~\cite{aubry1980}. For comparison, we will also study a random model in which the phase  is chosen randomly and independently on each site, $\phi_i^{R} =  \in [-\pi, \pi)$. We choose this form for the random fields instead of the more conventional uniform distribution~\cite{PalHuse,Luitz15} to keep the distribution of the on-site fields constant between the random and QP models, which enables a more direct comparison between the two. Both models are many-body localized for large field amplitudes $W > W_c^{QP/R}$.  We add the next-nearest neighbor terms with strength $J'$ to break the integrability of the models in the limit $W \rightarrow 0$, which allows the system to thermalize more completely within the thermal phase even for relatively small system sizes.

\begin{figure}
  \includegraphics[width=\columnwidth]{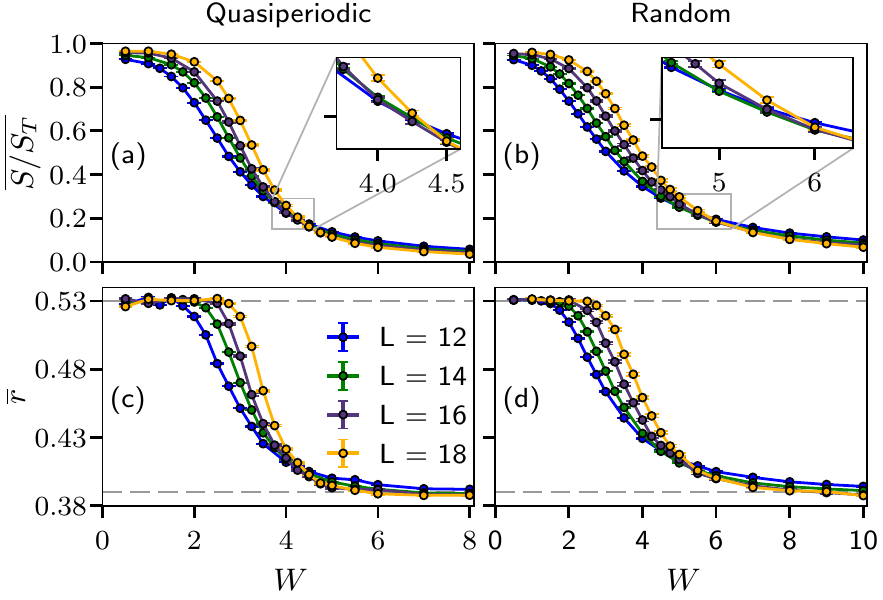}
  \caption{\label{fig:NNN}  (a), (b): Average half-chain eigenstate EE divided by the Page value $S_T$ for 
the quasiperiodic (a) and random (b) models. $S/S_T$ approaches a step function at the transition, going from zero in the MBL phase to one in the thermal phase. Insets show that the location of the crossings drift towards larger $W$ with increasing system size, but the finite-size drift is stronger in the random model. (c), (d): Level statistics ratio $\bar r$ which obeys GOE/Poisson distributions in the thermal/localized phases respectively in the quasiperiodic (c) and random (d) models. Both diagnostics show that the MBL phase in the quasiperiodic model is stable down to a lower value of $W$ as compared to the random one. }
\end{figure} 

Fig.~\ref{fig:NNN} benchmarks the location of the MBL transition(s) in \eqref{eq:model} using the half-chain entanglement entropy, $S$, and the level statistics ratio, $r$. Fig.~\ref{fig:NNN}(a)/(b) shows $S$ divided by 
$S_T = 0.5(L\log(2)-1)$, which is the Page~\cite{Page} value for a random pure state, in the quasiperiodic/random models respectively. The data is averaged over $1000-10^5$ disorder samples depending on $L$ (in the quasiperiodic model, the averaging is over different choices for the global phase shift $\phi^{QP}$), and over the middle quarter of the eigenstates in the $S^z_{\rm tot} = 0$ sector for each sample (for $L = 16,18$ we average over the middle 200 eigenstates). In both models, $S/S_T$ as a function of $W$ approaches a step function with increasing $L$, going from zero in the MBL phases with area-law entanglement to one in the thermal phase. 

Fig.~\ref{fig:NNN}(c)/(d) shows the level statistics ratio~\cite{OganesyanHuse} $ r \equiv \min\{\Delta_n, \Delta_{n+1}\}/\max\{\Delta_n, \Delta_{n+1}\}$, where $\Delta_n = E_n - E_{n+1}$ is the spacing between eigenenergy levels, in the quasiperiodic/random models respectively. This ratio approaches the GOE (Gaussian Orthogonal Ensemble) value $r \cong 0.53$ in the thermal phase and the Poisson value $r \cong 0.39$ in the localized phase for both models. 

A few points of note. 
First, the location of the crossing in the entropy/level statistics data drifts towards larger $W$ with increasing $L$ in both models, as is typical of all ED studies. However, the drifting of the crossing is stronger in the random model as compared to the QP one, suggesting that the QP model suffers less from this finite-size effect so the behavior we are seeing may be closer to the true asymptotic large-$L$ regime.  Second, as a related point, the transition is sharper (narrower in width) in the QP model. Third, despite the functional similarities in the choice of potentials between the two models, $  W_c^{QP} < W_c^{R}$, where we estimate $W_c^{QP} \gtrsim 4.25$  and $W_c^{R} \gtrsim 5.5$ (these are estimated as lower bounds since, as always, there is no observed crossover on the MBL side of the transition \cite{KhemaniCP}).  This means that the QP model remains localized down to a smaller value of $W$, which is most likely due to the absence of rare Griffiths events which can disrupt localization in the random model.  Indeed, within the MBL phase, the mean entanglement is larger in the random model than in the QP one (for comparable $W/W_c$), and distributions of the EE in the random model have longer tails to large entanglement reflecting rare events (see Supplement)

\begin{figure}
  \includegraphics[width=\columnwidth]{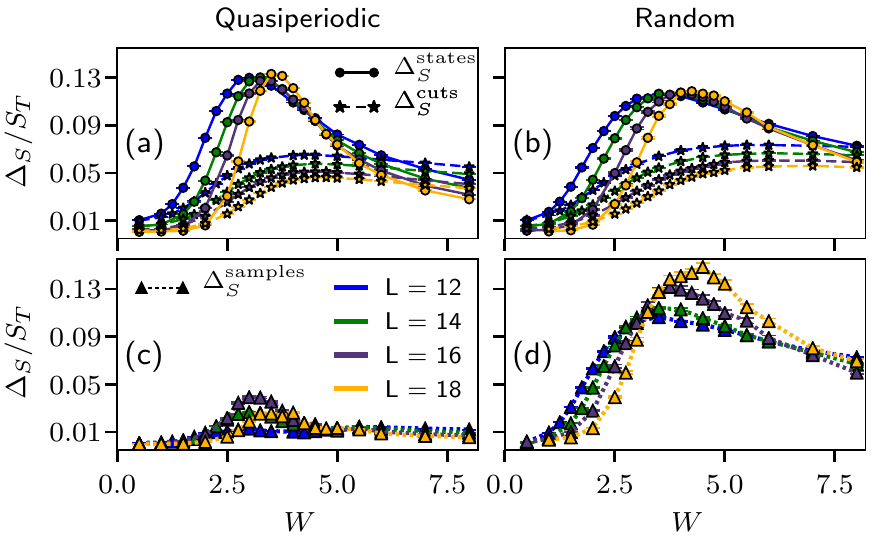}
  \caption{Standard deviation of the half-chain EE $\Delta_S$ divided by the Page value $S_T$, parsed by its contributions from eigenstate-to-eigenstate (solid, circles), cut-to-cut (dashed, stars),  and sample-to-sample (dotted, triangles) variations in the quasiperiodic (a,c) and random (b,d) models. The intra-sample variations look qualitatively similar between the two models (a,b) suggesting that these are mostly governed by the same fixed point at these sizes. However, the inter-sample variations are growing strongly with $L$ in the random model as it begins to crossover towards its asymptotic disorder dominated fixed point (d), while they are subdominant with no systematic $L$ dependence in the quasiperiodic model (c). }
\label{fig:Variance}
 \end{figure}

\noindent{\bf Variance of the half-chain entanglement entropy:}
We now study the standard deviation of the half-chain entanglement entropy $\Delta_S$ which peaks at the MBL transition, while it tends to zero deep in the MBL/ETH phases~\cite{Kjall14}. Following the prescription in Ref.~\onlinecite{KhemaniCP}, we parse the contributions to $\Delta_S$ due to fluctuations from sample-to-sample ($\Delta_S^{\rm samples}$)  from eigenstate-to-eigenstate within a given sample ($\Delta_S^{\rm states}$) and from different entanglement cuts within a given eigenstate ($\Delta_S^{\rm cuts}$);  see Fig~\ref{fig:Variance}. We use all cuts that produce a contiguous subsystem of length $L/2$. Since $S/S_T$ lies between $0$ and $1$, $\Delta_S/S_T$ can be at most 0.5.

First, note that the peak value of $\Delta_S^{\rm states}/S_T$ is independent of $L$ in both the QP (Fig.~\ref{fig:Variance}a) and random (Fig.~\ref{fig:Variance}b) models indicating a volume law scaling, $\Delta_S^{\rm states} \sim L$, in both and thus a substantial variance in $S$ across eigenstates of the \emph{same} sample. This property has not been included by any of the phenomenological RG approaches to the transition, and it indicates that the network of resonances driving the transition varies substantially across eigenstates of a given sample. Also note that the peak value of $\Delta_S^{\rm cuts}/S_T$ decreases with increasing $L$ (Figs.~\ref{fig:Variance}a, \ref{fig:Variance}b), indicating subvolume law scaling for $\Delta_S^{\rm cuts}$ in both models. This subvolume law scaling limits the spatial inhomogeneity of the resonant network of entanglement at the transition~\cite{KhemaniCP}.
Together, these data indicate that the \emph{intra}-sample critical variations across eigenstates and entanglement cuts look qualitatively similar between the random and QP models. In RG terms, this suggests that, for these sizes, the intra-sample finite-size critical behavior of the two models are perhaps governed mostly by the same fixed point (c.f. Fig.~\ref{fig:RG}).

On the other hand, the two models look strikingly different when considering \emph{inter}-sample variations.  In the quasiperiodic model, $\Delta_S^{\rm samples}$ is far sub-dominant to the intra-sample contributions and is not growing systematically with $L$ (Fig.~\ref{fig:Variance}c).  This indicates that the different quasiperiodic samples are quantitatively similar in their entanglement properties, and sample-to-sample fluctuations are \emph{not} the dominant source of the finite-size critical rounding in the quasiperiodic model at these sizes.  

By contrast, in the random model, the peak value of $\Delta_S^{\rm samples}/S_T$ grows strongly with $L$ which naively indicates that  $\Delta_S^{\rm samples}$ scales super-linearly with $L$ (Fig.~\ref{fig:Variance}d), a trend that is not sustainable in the asymptotic large $L$ limit. This indicates that effects of inter-sample quenched randomness are not yet fully manifest but growing strongly at these small sizes. In RG terms, we interpret this as an indication of an RG flow, due to the external randomness, that is away from the fixed point that governs the nonrandom quasiperiodic model and is towards the infinite-randomness Harris/Chayes obeying fixed point that will asymptotically govern the transition for this random model  (c.f. Fig.~\ref{fig:RG}).

\begin{figure}[t]
  \includegraphics[width=\columnwidth]{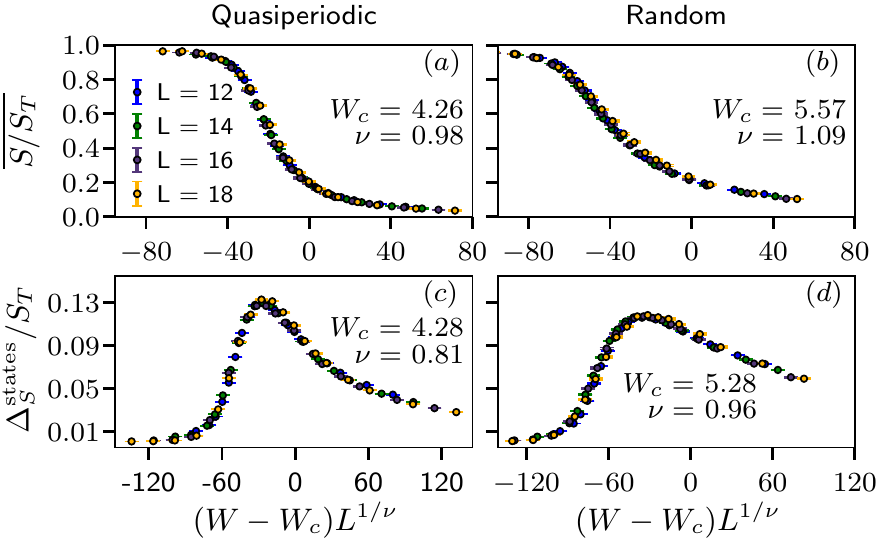}
  \caption{ Finite-size critical scaling collapse for $S$ (a,b), and $\Delta_S^{\rm states}$ (c, d) data in the quasiperiodic and random models. We see that $\nu \sim 1$ for both models, again suggesting that the transition in both models is mostly governed by the same nonrandom fixed point at these sizes. This exponent is in violation of CCFS/CLO bounds which must asymptotically constrain the random model - note that $\nu$ is slightly larger for the random model consistent with the suggestion that the effects of quenched randomness are growing but not yet fully apparent at these sizes. The critical $W_c$ is larger in the random model.    }
  \label{fig:ScalingCollapse} 
\end{figure}

\noindent{\bf Two universality classes:} We now turn to the finite-size critical scaling properties of the MBL transition in the two models. Figure~\ref{fig:ScalingCollapse} shows scaling collapse for $S/S_T$ and $\Delta_S^{\rm states}/S_T$, where both quantities are fit to a form $g[(W - W_c)L^{1/\nu}]$, where $W_c$ denotes the critical disorder strength and $\nu$ is the finite-size scaling exponent. 

We see a scaling collapse in the quasiperiodic model with $W_c \sim 4.25$ and $\nu \sim 1$ (Fig~\ref{fig:ScalingCollapse} a,c). First, note that quasiperiodic models without quenched randomness are not subject to the CCFS/CLO bound which require $\nu \geq 2/d$. Instead, such models fall under the purview of the Harris-Luck criterion~\cite{Luck} which imposes the weaker bound $\nu \geq 1/d$~\footnote{ Luck generalizes the Harris bound on the relevance of disorder near a clean equilibrium phase transition to the more general aperiodic case \cite{Luck}. This work, however, still refers to a clean critical point and the equivalent of the CCFS \cite{CCFS2} scaling bounds which do not refer to a clean transition do not exist for the general aperiodic case to the best of our knowledge.}. The observed scaling exponents are certainly already quite close to obeying this bound, considering the small sizes studied.  This, combined with our observations of finite-size drifts in the discussion surrounding Fig~\ref{fig:NNN}, suggests that the critical behavior in the quasiperiodic model might be close to its asymptotic large-$L$ form even at these sizes. 
If the scaling exponent continues to be $\nu \sim1 $ even in the asymptotic limit, then it is clear that the MBL transition in quasiperiodic models belongs to a different universality class from the transition in models with quenched randomness which must obey the CCFS/CLO bound---this would make the external randomness Harris-relevant when added to the quasiperiodic model (c.f. Fig~\ref{fig:RG}).  

It is an interesting curiosity that the non-interacting Aubry-Andre transition  also has $\nu =1$, so one might be tempted to believe that the critical properties of the interacting quasiperiodic transition belong to the same universality class as the non-interacting one. However, a careful analysis (not shown) reveals that properties like the volume law scaling of $\Delta_S^{\rm states}$ across the many-body eigenstates are absent in the non-interacting model. 

Turning to the random model, we see a scaling collapse with a larger critical disorder strength $W_c \sim 5.5$  (Fig~\ref{fig:ScalingCollapse} b,d) which is consistent with the presence of rare Griffiths effects in the random model which can aid with thermalization. The scaling exponent $\nu \sim 1$ confirms our earlier observation that the transition in the random model looks in many respects like it belongs to the quasiperiodic universality class at these sizes which are too small to feel the full effects of the quenched randomness. Also note that the scaling exponent is consistently slightly larger for the random model as compared to the quasiperiodic one, which is congruent with the theory that the random model is ``en-route'' to crossing over to a different disorder dominated scaling regime with $\nu \geq 2$ at larger system sizes.  

\noindent{\bf Summary and outlook:} We systematically examined the MBL transition in random and quasiperiodic models, and found that the  MBL phase is stable down to a smaller disorder strength in the quasiperiodic case.  Moreover, finite-size scaling analysis near the transition strongly suggests that the quasiperiodic model asymptotically belongs to a different universality class from the random one. We find scaling exponents $\nu \sim 1$ for both models; however while this exponent may be close to its asymptotic value for the quasiperiodic model (and in agreement with the Harris-Luck bound),  we know that the asymptotic scaling exponent in the disordered model must satisfy $\nu \geq 2/d$ because the width of the finite-size scaling window is  constrained to be greater than $\sim L^{-d/2}$ due to sample-to-sample fluctuations from the quenched randomness. Indeed, the sample-to-sample standard deviation of the entanglement entropy in the random model clearly shows that the effects of randomness are not fully apparent, but growing strongly, at the sizes studied, and many critical properties of the random models at these sizes look similar to those of quasiperiodic models. In RG terms, the transition in both the random and quasiperiodic models appears to be governed by the same nonrandom fixed point for the sizes accessible to ED, but the random model is starting to crossover towards the disorder dominated fixed point. 

Additionally, the entanglement structure at the critical fixed points in RG studies \cite{VHA, PVP} indicates that the asymptotic disorder-dominated regime in these random models might only be apparent in samples larger than $\sim$ 100 spins ~\cite{KhemaniCP}, which will most likely remain inaccessible to both experimental and numerical work. Our work indicates that there should be a greater focus on quasiperiodic models in finite-size studies of the MBL transition, since the asymptotic scaling regime of the transition is likely more accessible in such models. Further, it is possible that the MBL phase in quasiperiodic models is more stable even in higher dimensions and for longer-ranged interactions since the recent arguments~\cite{de2016stability} on the instability of MBL due to rare, thermal inclusions arising from disorder fluctuations don't apply to quasiperiodic models. Of course, a renormalization group study of the transition in a quasiperiodic model, if possible, would be a helpful next step for better understanding the properties of this new universality class.  It is also intriguing to ask whether the two cases studied in the present work  cover all universal possibilities for MBL transitions, or if there are further classifications --- say for example in the case of a transition to an MBL phase accompanied by the simultaneous development of spontaneous symmetry breaking \cite{Huse13, PekkerHilbertGlass, VasseurParticleHole}, or for MBL transitions in models with correlated disorder with varying degrees of correlation. 

{\it Acknowledgements:}
We thank Anushya Chandran, Chris Laumann, Subroto Mukerjee, Siddharth Parameswaran, Andrew Potter and Shivaji Sondhi for stimulating discussions, and Liangsheng Zhang for preliminary work on related models.  This work was supported by the Harvard Society of Fellows (VK) and NSF grant DMR-1408560 (DS).

\bibliography{global}

\end{document}


\title{Supplement: Two universality classes for the many-body localization transition}

\author{Vedika Khemani}
\affiliation{\mbox{Department of Physics, Harvard University, Cambridge, MA 02138, USA}}
\author{D. N. Sheng}
\affiliation{\mbox{Department of Physics and Astronomy, California State University, Northridge, CA 91330, USA}}
\author{David A. Huse}
\affiliation{\mbox{Department of Physics, Princeton University, Princeton, NJ 08544, USA}}
\maketitle

\section{ Distributions of Entanglement Entropy}

In this supplement, we present data for the distributions of the half-chain entanglement entropy in the random and quasiperiodic models.  These carry more information than the mean values presented in Fig.~1 in the main text, and provide evidence that the random models suffer more from rare ``large entanglement'' Griffiths effects which can disrupt localization and lead to a lower relative stability of the MBL phase in the random model. 

 \begin{figure} [h]
  \includegraphics[width=0.5\columnwidth]{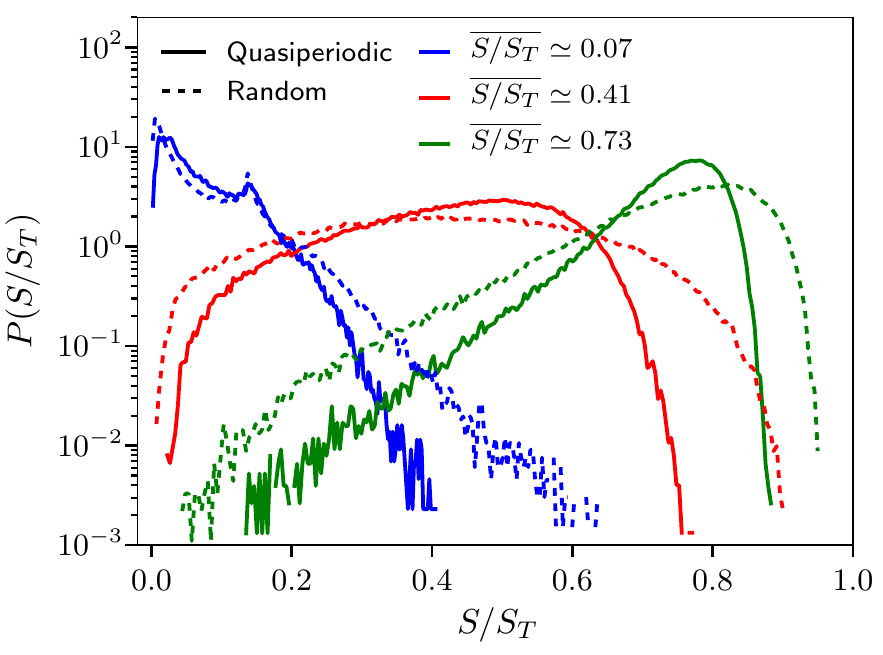}
  \caption{Probability distributions of the normalized mid-cut entanglement entropy $S/S_T$ corresponding to three different mean values of $S/S_T$ (blue, red and green) for the quasiperiodic (solid lines) and random (dashed line) models in a system of size $L=16$. The distributions in the random model consistently have a longer tail out to large entanglement.  } 
  \label{fig:EntropyDist}  
 \end{figure}

Fig.~\ref{fig:EntropyDist} shows distributions of the mid-cut entanglement entropy (normalized by the Page value) corresponding to three different mean values of $S/S_T$ for the random and quasiperiodic models in a system of size $L=16$. Due to the differences in the location of the transition in these models, the chosen mean entanglement values correspond to different $W$'s in the two models, and keeping the mean fixed  permits a more sensible comparison of the distributions in the two cases. We see that, for the same mean value of the entanglement entropy, the distributions in the random model are broader, with longer tails to large and to small entanglement, reflecting the presence of rare Griffiths regions in the random model.  This is true in all cases shown: on the MBL side of the MBL-to-thermal crossover ($\overline{S/S_T} \simeq 0.07$), near the MBL-to-thermal crossover  ($\overline{S/S_T} \simeq 0.41$) and on the thermal side of the crossover ($\overline{S/S_T} \simeq 0.73$).